\documentclass[utf8]{FrontiersinHarvard} 

\usepackage{url,hyperref,lineno,microtype,subcaption}
\usepackage[onehalfspacing]{setspace}

\usepackage{etoolbox}
\usepackage{graphicx}
\usepackage[flushleft]{threeparttable}

\usepackage{url}
\usepackage{hyperref}

\hypersetup{breaklinks=true,colorlinks=true,linkcolor=blue,citecolor=blue,filecolor=blue,urlcolor=blue}
\makeatletter
\patchcmd{\NAT@citex}
  {\@citea\NAT@hyper@{%
     \NAT@nmfmt{\NAT@nm}%
     \hyper@natlinkbreak{\NAT@aysep\NAT@spacechar}{\@citeb\@extra@b@citeb}%
     \NAT@date}}
  {\@citea\NAT@nmfmt{\NAT@nm}%
   \NAT@aysep\NAT@spacechar\NAT@hyper@{\NAT@date}}{}{}

\patchcmd{\NAT@citex}
  {\@citea\NAT@hyper@{%
     \NAT@nmfmt{\NAT@nm}%
     \hyper@natlinkbreak{\NAT@spacechar\NAT@@open\if*#1*\else#1\NAT@spacechar\fi}%
       {\@citeb\@extra@b@citeb}%
     \NAT@date}}
  {\@citea\NAT@nmfmt{\NAT@nm}%
   \NAT@spacechar\NAT@@open\if*#1*\else#1\NAT@spacechar\fi\NAT@hyper@{\NAT@date}}
  {}{}

\def\aj{Astron.~J.}
\def\araa{Ann.~Rev.~Astron.~Astrophys.}
\def\apj{Astrophys.~J.}
\def\apjl{Astrophys.~J.~Lett.}
\def\apjs{Astrophys.~J.~Suppl.}

\def\aap{Astron.~Astrophys.}

\def\aaps{Astron.~Astrophys.~Suppl.~Ser.}

\def\mnras{Mon.~Not.~R.~Astron.~Soc.}

\def\prd{Phys.~Rev.~D}

\def\prl{Phys.~Rev.~Lett.}

\def\pasp{Publ.~Astron.~Soc.~Pac.}

\def\ssr{Space~Sci.~Rev.}

\def\nat{Nature}

\def\physrep{Phys.~Rep.}

\usepackage{listings}

\definecolor{dkgreen}{rgb}{0,0.6,0}
\definecolor{gray}{rgb}{0.5,0.5,0.5}
\definecolor{mauve}{rgb}{0.58,0,0.82}

\def\keyFont{\fontsize{8}{11}\helveticabold }
\def\firstAuthorLast{Danehkar} 
\def\Authors{A. Danehkar\,$^{1,*}$}
\def\Address{$^{1}$Eureka Scientific, 2452 Delmer Street, Suite 100, Oakland, CA 94602-3017, USA}
\def\corrAuthor{Corresponding Author}

\def\corrEmail{danehkar@eurekasci.com}

\begin{document}
\onecolumn
\firstpage{1}

\title[Relativistic Reflection and Related Variability]{Relativistic Reflection Modeling in AGN and Related Variability from PCA: A Brief Review} 

\author[\firstAuthorLast ]{\Authors} 
\address{} 
\correspondance{} 

\extraAuth{}

\maketitle


\begin{abstract}
\noindent X-ray observations of active galactic nuclei (AGNs) reveal relativistic reflections from the innermost regions of accretion disks, which contain general-relativistic footprints caused by spinning supermassive black holes (SMBH). We anticipate the spin of a SMBH to be stable over the human timeframe, so brightness changes in the high-energy corona above the SMBH should slightly alter relativistic reflection. In this brief review, we discuss the latest developments in modeling relativistic reflection, as well as the rapid small variation in relativistic emission disclosed by the principal component analysis (PCA) of X-ray variability in AGN. PCA studies of X-ray spectra from AGNs have shown that relativistically blurred reflection has negligible fluctuations over the course of observations, which could originate from rapid (intrahour) intrinsic variations in near-horizon accretion flows and photon rings. The PCA technique is an effective way to disclose relativistic reflection from X-ray observations of AGNs, simplifying the complexity of largely variable X-ray data for automated spectral analysis with machine learning algorithms.
\tiny
 \keyFont{ \section{Keywords:} active galactic nuclei, relativistic disks, black hole spin, reflection, X-ray sources, principal component analysis} 
\end{abstract}


\section{Introduction}
\label{review:introduction}


The center of of the Milky Way is characterized by a supermassive black hole (SMBH), which is supported by indirect but compelling observational evidence such as stellar orbits in the vicinity of Sagittarius A* \citep[Sgr\,A*;][]{Ghez1998,Ghez2005} and the near-infrared luminosity of Sgr\,A* being consistent with the presence of an event horizon \citep{Broderick2006,Broderick2009}. Similarly, we expect that active galactic nuclei (AGNs) in other galaxies host SMBHs at their centers \citep{Kormendy1988,Kormendy1992,Kormendy1997,Cretton1999}, which are essential to explaining the X-ray features of quasars and AGNs \citep[see review by][]{Mushotzky1993}. Several techniques, such as the reverberation mapping \citep{Blandford1982}, spectral energy distribution (SED) fitting \citep{Shields1978,Malkan1983}, and broad-line region size--luminosity correlation \citep{Vestergaard2002}, have been developed to validate the presence of SMBHs and estimate their masses \citep[e.g.,][]{Kormendy1995,Miyoshi1995,Wandel1999,Peterson2004,Calderone2013,Capellupo2015,Bentz2015,Mejia-Restrepo2016}. Our constraints on SMBH masses have allowed us to establish the connections between SMBHs and the evolution of their host galaxies \citep[e.g.,][]{Magorrian1998,Ferrarese2000,Haering2004,Heckman2014}. 

Some solutions of standard general relativity simply characterize black holes using two parameters, mass and spin \citep{Kerr1963}, which can fully describe the properties of SMBHs. In this regard, spins of SMBHs, along with masses, could produce some of the fundamental mechanisms for powering relativistic jets \citep[e.g.,][]{Garofalo2010,Tchekhovskoy2012}, as well as describing the discrepancy between radio-loud and radio-quiet AGNs \citep{Wilson1995,Moderski1998}, galaxy evolution \citep{DiMatteo2005,Volonteri2013,Sesana2014}, and galaxy mergers \citep{Hughes2003,Volonteri2005,Berti2008}.
In particular, ultra-fast outflows (UFOs) have been detected in X-ray observations of several radio-quiet AGNs \citep[e.g.,][]{Tombesi2010,Tombesi2011,Tombesi2012,Danehkar2018,Boissay-Malaquin2019}, while extended relativistic jets have been seen in radio observations of radio-loud AGNs \citep[see review by][]{Blandford2019}.
The spins of SMBHs could have a potential role in the formation of UFOs and jets seen in AGNs and quasars \citep{MacDonald1986,Thorne1986}.  
These phenomena can be explained by spinning SMBHs according to the Blandford--Znajek \citep{Blandford1977} and Penrose mechanism \citep{Penrose1969,Penrose2002,Penrose1971}, as well as frame-dragging vortexes \citep[e.g.,][]{Owen2011,Nichols2011,Danehkar2020}. Alternatively, they could originate magnetically from the innermost accretion disk in the vicinity of a spinning SMBH according to the Blandford--Payne mechanism \citep{Blandford1982a}.

In the Boyer--Lindquist coordinates, the Kerr metric \citep{Kerr1963} of a spinning black hole is expressed using the set of oblate spheroidal coordinates ($r$, $\theta$, $\phi$) as follows \citep{Boyer1967}:
\begin{align}
  ds^2=&-\left(1-\frac{r_{\rm s} r}{\Sigma}\right)c^2 dt^2 -
  \frac{2\bar{a} r_{\rm s} r\sin^2\theta}{\Sigma}c dt\,d\phi +
  \frac{\Sigma}{\Delta}dr^2 \nonumber \\
&+\Sigma\,d\theta^2+\left(r^2+\bar{a}^2+\frac{\bar{a}^2 r_{\rm s} r\sin^2\theta}{\Sigma}\right)\sin^2\theta\,d\phi^2, \label{eq_1}
\end{align}
where $\Sigma = r^2+\bar{a}^2\cos^2\theta$, $\Delta=r^2-r_{\rm s}r+\bar{a}^2$, $r_{\rm s}=2G M/c^2$ is the Schwarzschild radius, $\bar{a}=J/Mc$ is called the Kerr spin parameter describing angular momentum per unit mass having the length dimension, $J$ the black hole angular momentum, $M$ the black hole mass, $G$ the Newtonian constant of gravitation, and $c$ the speed of light. The \textit{dimensionless spin parameter}, which is frequently used in the astrophysical community, is defined as $a\equiv 2\bar{a}/r_{\rm s}=Jc/(GM^2)$, while $-1\leq a \leq 1$; negative values describe retrograde rotation, in which the black hole rotates in the opposite direction of the accretion disk, whereas positive values are associated with prograde rotation, and zero implies non-rotating black holes. The outer and inner event horizons are determined by the roots of $\Delta=0$, which are $r_{\pm}=r_{\rm s}(1\pm\sqrt{1-a^2})/2$. In the case of $a=0$, Eq. (\ref{eq_1}) reduces to the Schwarzschild metric \citep{Schwarzschild1916} with the event horizon at $r=r_{\rm s}$. Unlike the Schwarzschild metric, which has a singularity at $r = 0$, the Kerr metric has one at $\Sigma = 0$.
The innermost stable circular orbit (ISCO) of the accretion disk is located at a radius of marginal stability, $r_{\rm ms}$, which is given by \citep{Bardeen1972}:
\begin{align}
r_{\rm ms} (a)=&\frac{GM}{c^2}\left(3+Z_2 -{\rm sgn}(a) \sqrt{(3-Z_1)(3+Z_1+2Z_2)} \right),
\label{eq_2}
\end{align}
where ${\rm sgn}(a)$ is the signum function having the value $-1$, $1$ or $0$ according to the sign of $a$, while $Z_1$ and $Z_2$ are defined as,
\begin{align}
Z_1=&1+\left(1-a^2\right)^{1/3}\left[\left(1+a\right)^{1/3}+\left(1-a\right)^{1/3}\right],\\
Z_2=&\sqrt{3a^2+Z_1^2}.
\end{align}
This implies that the accretion disk has a limited extent at the marginal stability radius ($r_{\rm ms}$), also called the ISCO radius. The value of this radius depends upon the dimensionless spin parameter, e.g.,
\begin{equation}
  r_{\rm ms} =
    \begin{cases}
      0.5 r_{\rm s} & \text{for $a=1$}\\
      3 r_{\rm s} & \text{for $a=0$}\\
      4.5 r_{\rm s} & \text{for $a=-1$}
    \end{cases}.  
\label{eq_3}
\end{equation}
In prograde rotation, the ISCO radius shrinks to nearly half of the Schwarzschild radius as it approaches a near-maximal spin ($a \approx 1$), while it expands in retrograde rotation ($-1<a<0$). For a non-spinning black hole, the ISCO radius is precisely three times the Schwarzschild radius.

There are different methods available to measure the spin of a single SMBH \citep[see review by][]{Brenneman2013}. All of them are based on general relativity solutions of the Kerr spacetime in the vicinity of the black hole. They use the aforementioned fact that the ISCO radius ($r_{\rm ms}$) of the accretion disk depends on the spin, as seen in Eq. (\ref{eq_2}), and assume a geometrically thin disk that is capable of irradiating the corona radiation with light-bending in the innermost regions (see Fig.~\ref{fig:1} bottom-left). Below are the techniques that have been used to constrain the spins of SMBHs:
\begin{itemize}
\item \textbf{X-ray Reflection Spectroscopy.} High-energy radiation from a corona or the base of a jet illuminates the accretion disk, reflecting scattered photons, which forms the basis of this method. Multiple Compton scatterings (Comptonization) of soft thermal photons lead to the cooling of the hot electrons in the corona \citep{Haardt1991,Haardt1993a}. A portion of the comptonized radiation undergoes scattering outside of the ionizing source, resulting in the formation of a power-law-shaped continuum that is typically observed in X-rays from AGN \citep{Haardt1991}. However, a fraction of the scattered photons will undergo reflection on the surface of the disk \citep{Haardt1993a}, as seen in Fig.~\ref{fig:1} (top). If the disk is not fully ionized, the continuum includes the emission of various fluorescent emission lines at energies below 7\,keV, in addition to the Compton hump with a peak at around 20--30\,keV caused by downscattering, as seen in Fig.~\ref{fig:1} (top panel). The most notable line is Fe K$\alpha$, with a rest-frame energy of 6.4\,keV, which is produced due to the significant iron abundance and fluorescence process. This line is the most important tool for describing the relativistic reflection from the innermost disk, as it becomes broadened and skewed due to Doppler and general-relativistic effects \citep[see reviews by][]{Reynolds2003,Miller2007,Reynolds2013,Reynolds2014,Reynolds2019,Bambi2021}. The truncation of its low-energy tail directly corresponds to the ISCO radius, i.e., the spin. This feature, independent of mass or distance from the black hole, enables the measurement of black hole spins. One of the drawbacks of this method for AGN is the complex absorption from line-of-sight material, typically found at lower energies in the red tail. Moreover, exceptionally high counts are required to properly constrain the spin; $2 \times 10^{5}$ \citep{Guainazzi2006} or $1.5 \times 10^{5}$ counts \citep{deLaCallePerez2010} in the energy range of 2--10 keV. Furthermore, this approach is actually model-dependent, as demonstrated by a list of reflection models in Table~\ref{ref:rel:model}. However, the advancements in X-ray spectroscopy above 10\,keV with the \textit{Nuclear Spectroscopic Telescope Array} \citep[\textit{NuSTAR};][]{Harrison2013}, which has no pile-up effects, have significantly improved the reliability of this approach by including the Compton hump reflection at high energies \citep[$> 10$ keV, peaking within $20$--30 keV; see, e.g.,][]{Parker2014b,Keck2015,Victoria-Ceballos2023}. Most of the spectral models developed for X-ray relativistic reflection are briefly reviewed in Sec.\,\ref{review:model}.

\begin{figure}[h!]
\begin{center}
\includegraphics[width=0.95\textwidth, trim = 0 0 0 0, clip, angle=0]{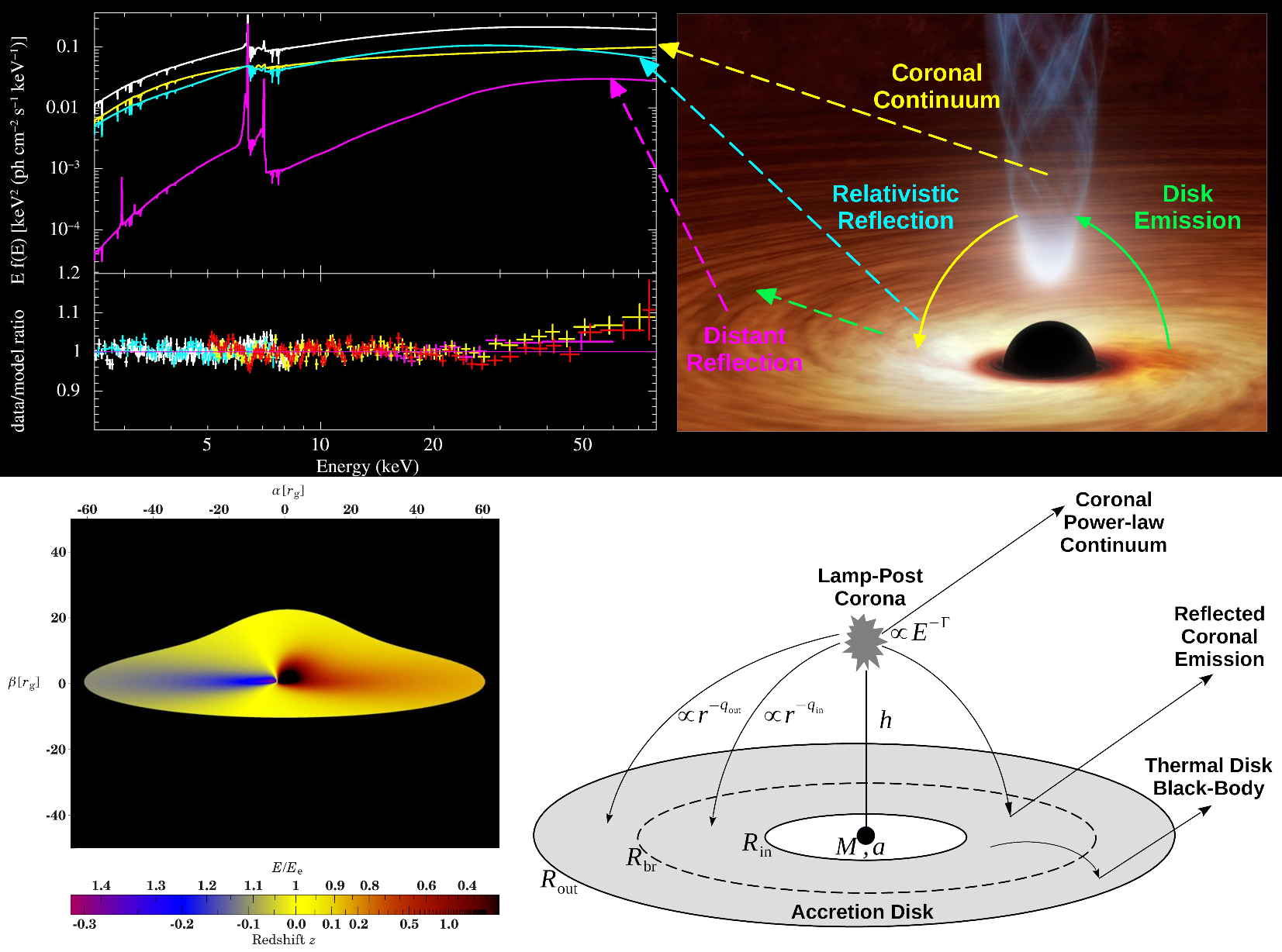}%
\end{center}
\caption{Top: A schematic view of X-ray relativistic reflection. Spectral model (left panel) of the  Seyfert 1.5 galaxy NGC\,4151, consisting of a coronal continuum (\textsf{highecut}$\times$\textsf{zpowerlw}; yellow), relativistic reflection (\textsf{relconv}$\times$\textsf{xillver}; light blue), and distant reflection (\textsf{xillver}; purple), fitted to \textit{NuSTAR} and \textit{Suzaku} observations, from \citet{Keck2015}. An artist's illustration (right) of the radiation reflection from the accretion disk around a black hole (courtesy of NASA/JPL-Caltech/R.\,Hurt at IPAC/R.\,Connors at Caltech).
Bottom: A simulated image (left panel) of an accretion disk, inclined with $i = 80^{\circ}$ (relative to the line of sight), gravitationally distorted by general-relativistic light-bending effects of a black hole at the center, from \citet{Garcia2014}. A diagram (right) illustrating the lamp-post corona with a power-law-shaped continuum ($\propto E^{-\Gamma}$) situated at a height ($h$) above a black hole simply described by its mass ($M$) and spin ($a$), and the innermost accretion disk characterized by the boundary radii ($R_{\rm in}$ and $R_{\rm out}$) and the break radius ($R_{\rm br}$) to describe the coronal emissivity laws $\propto r^{-q_{\rm in}}$ ($R_{\rm in}<r<R_{\rm br}$) and $\propto r^{-q_{\rm out}}$ ($R_{\rm br} < r< R_{\rm out}$), as well as corresponding thermal disk black-body and reflected coronal emission, adopted from \citet{Dauser2014a} and \citet{Hoormann2016}.
\label{fig:1}
}
\end{figure}

\item \textbf{Broad-band SED Fitting.} This method was initially began to be deployed for X-ray binaries by \citet{Zhang1997} and \citet{Gierlinski2001}. This approach depends on the distance, mass, and disk inclination angle of the accretion disk \citep[see review by][]{Remillard2006}, so it has mostly been used for the spin measurement of stellar-mass black holes \citep[e.g.,][]{Shafee2006,McClintock2006}. This method was first exploited by \citet{Done2013} to put constraints on the spins of SMBHs with the \textsf{optxconv} model \citep[based on \textsf{optxagnf};][]{Done2012}, which contains the SED spectrum made by a (color-temperature-corrected) blackbody, an optically thick warm Comptonisation (soft excess; $<2$\,keV) component, and an optically thin, hot Comptonisation (power-law; $>2$\,keV) component. Later, the same technique was employed to measure the spins of AGNs at $z \sim 1.5$ that evolved just after cosmic noon \citep{Capellupo2015,Capellupo2016}, and was benchmarked against X-ray reflection measurements for NGC 3783, an AGN known for its relativistically broadened Fe K$\alpha$ line \citep{Capellupo2017}. 
This method was employed to constrain the masses and spins of SMBSH in four blazars at high redshifts \citep{Campitiello2018}, which was implemented using the \textsf{kerrbb} model \citep{Li2005} of multi-temperature blackbody spectrum of a thin accretion disk around a spinning black hole. 
The \textsf{optxconv} SED model \citep{Done2012,Done2013} was also used by \citet{Porquet2019} to derive a well-measured spin rate of the SMBH in Ark\,120, a well-known bare AGN with no intrinsic absorption along the line-of-sight. Subsequently, new SED models of the broad-band continuum of AGN, called  \textsf{agnsed} and \textsf{qsosed} \citep[][]{Kubota2018,Petrucci2018}, have been developed, followed by a super-Eddington accretion model of the slim disk \citep[\textsf{agnslim};][]{Kubota2019}, which provide better constraints on the masses and spins of SMBHs.
More recently, \citet{Hagen2023} made a fully general-relativistic implementation of \textsf{agnsed}, referred to
as the \textsf{relagn} model that includes general-relativistic ray tracing and the relativistic Novikov--Thorne disk model \citep{Novikov1973}, leading to a complex disk spectrum in the soft excess instead of a simple blackbody, and was utilized to determine the SMBH spin in Fairall\,9 from the broad-band spectrum, extending from Optical/UV to the X-ray.
 
\item \textbf{Radio Event Horizon Imaging.} This method employs sub-mm data collected by several very long baseline interferometry (VLBI) stations over different locations (e.g., JCMT, SMT, SPT, IRAM, APEX, and ALMA) to achieve micro-arcsecond spatial resolution images of an SMBH event horizon \citep{EventHorizonTelescopeCollaboration2019,EventHorizonTelescopeCollaboration2022}. This has enabled the first-ever images of the accretion flow in the vicinity of nearby SMBHs to be produced, namely M87 and Sgr\,A* \citep{EventHorizonTelescopeCollaboration2019a,EventHorizonTelescopeCollaboration2022a}. We can determine the SMBH spin by accurately modeling the appearance of accretion flows in VLBI images, accounting for general-relativistic light bending (see Fig.~\ref{fig:1} bottom-left) based on various characteristics such as the ISCO radius ($r_{\rm ms}$). The VLBI imaging techniques have recently been used to deduce the spin values according to high-spatial-resolution images in Sgr\,A* and M87 \citep{EventHorizonTelescopeCollaboration2019b,EventHorizonTelescopeCollaboration2021,EventHorizonTelescopeCollaboration2022b,EventHorizonTelescopeCollaboration2022c}. This method has the disadvantage of only being suitable for nearby SMBHs.

\end{itemize}

\section{Relativistic Reflection Modeling}
\label{review:model}

Various spectral models have been constructed to reproduce the general-relativistic effects of the Kerr metric on the iron K$\alpha$ line profile. Table~\ref{ref:rel:model} summarizes most of the well-known spectral models made for X-ray data analysis of relativistically blurred emission, along with their key parameters. The basic parameters in these models are the dimensionless spin parameter ($a=Jc/GM^2$), the line-of-sight inclination angle ($i$) of the accretion disk, the spectral photon index ($\Gamma$) of the primary source -- the corona or the base of a jet above the black hole --, as well as the boundary radii ($R_{\rm in}$ and $R_{\rm out}$) of the innermost accretion disk. Most of these models can be loaded into tools for X-ray spectral analysis, particularly the X-ray spectral fitting package \textsc{xspec}\footnote{\url{https://heasarc.gsfc.nasa.gov/xanadu/xspec/}} \citep{Arnaud1996,Arnaud1999} of HEASoft's data analysis package for X-ray astronomy \textsc{xanadu} \citep{NHEASARC2014}, MIT's Interactive Spectral Interpretation System\footnote{\url{https://space.mit.edu/cxc/isis/}} \citep[\textsc{isis};][]{Houck2000}, CXC\footnote{The Chandra X-Ray Center (CXC) is operated for NASA by the Smithsonian Astrophysical Observatory (SAO).}'s Modeling and Fitting Package \textsf{Sherpa}\footnote{\url{https://cxc.cfa.harvard.edu/sherpa/}} \citep{Freeman2001,Doe2007} of the Chandra Interactive Analysis of Observation \citep[\textsc{ciao};][]{Fruscione2006}, and SRON\footnote{Netherlands Institute for Space Research (Stichting Ruimteonderzoek Nederland; SRON) is a Dutch institute for astrophysical research.}'s X-ray high-resolution spectral modeling and fitting package \textsc{spex}\footnote{\url{https://www.sron.nl/astrophysics-spex/}} \citep{Kaastra1996,dePlaa2020}. These models are often applied to integrated spectra of AGNs without accounting for the variability of the X-ray sources.

The early models, developed to analyze relativistic reflection, featured fiducial values for the spin parameter. 
Black hole spin measurement began with the \textsf{diskline} \citep{Fabian1989} and \textsf{laor} \citep{Laor1991} fixed-spin models, which were run with fiducial spin values of $a = 0$ and $0.998$, respectively. The \textsf{diskline} model was based on analytic, time-consuming calculations, whereas the \textsf{laor} model relied on extensive pre-calculated tabulated Flexible Image Transport System (FITS) data created for different combinations of the model parameters: inclination angle, spectral photon index ($\Gamma$), inner radius ($R_{\rm in}$), and outer radius ($R_{\rm out}$). The \textsf{laor} model recreates the relativistic line shape by interpolating data from the extensive FITS table. Later, \citet{Martocchia1996} examined how the `lamp post' geometry affected the broad iron K$\alpha$ lines, in which the ionizing source is located on the polar axis at a height ($h$) above the black hole, as illustrated in Fig.~\ref{fig:1} (bottom-right). This investigation was followed by a comprehensive analysis for $a=0.001$ and $0.9981$ leading to the \textsf{kerrspec} model \citep{Martocchia2000,Martocchia2002}. We should note that the reflection continuum was not included in these models and needed to be handled by a separate model such as \textsf{reflionx} \citep{Ross1999,Ross2005,Ross2007}, which should be convolved with the corresponding convolution models to make a smoothed spectrum of relativistic smearing in an accretion disk. The convolution models \textsf{rdblur} and \textsf{kdblur} were prepared using \textsf{diskline} and \textsf{laor}, respectively, which can be used with a reflection model (e.g., \textsf{reflionx}). \citet{Nandra2007} used a modified version of the \textsf{kdblur} model to characterize the broad iron K$\alpha$ line in \textit{XMM-Newton} observations of a sample of Seyfert galaxies. However, the spectral models with fixed spin rates obviously prevented us from straightforwardly measuring the black hole spin.

The next generation of relativistic reflection models has a free parameter for the positive spin rates, which allows for the determination of the black hole spin in prograde rotation ($0 \leqslant a < 1$). The \textsf{kyrline} (or \textsf{ky}) model \citep{Dovciak2004a,Dovciak2004,Dovciak2022} was developed to incorporate extensive tables calculated for transfer functions. This model rapidly computes the shape of relativistically broadened line emission using transfer function tables without relying heavily on interpolation, yielding a significantly greater level of spectral resolution compared to the \textsf{laor} model. Furthermore, the black hole spin was a free parameter, ranging from $a=0$ to $1$. The model can therefore reproduce the relativistic blurred emission more accurately than the \textsf{laor} model for all positive spin rates and inclination degrees. In addition, the \textsf{kyrline} model features coronal emissivity law indexes ($q_{\rm in}$ as $r^{-q_{\rm in}}$ between $R_{\rm in}$ and $R_{\rm br}$; and $q_{\rm out}$ as $r^{-q_{\rm out}}$ between $R_{\rm br}$ and $R_{\rm out}$) for precisely creating the emitted radiation (see Fig.~\ref{fig:1} bottom-right), describing the emissivity characteristics on both sides of the break radius ($R_{\rm br}$), as well as including the \textsf{limb} prescriptions for limb-darkening/-brightening laws \citep{Chandrasekhar1960}, namely isotropic emission ($I \propto 1$), Laor's limb-darkening \citep[$I \propto 1+2.06\mu$;][]{Laor1991}, and Haardt's limb-brightening \citep[$I \propto \ln (1+1/\mu)$;][]{Haardt1993}, where $\mu=\cos(\theta_{\rm e})$ and $\theta_{\rm e}$ is the inclination angle of the emitted radiation with respect to the disk. These features were accomplished by computing them for inclusion in the comprehensive FITS table. In order to address the problems with the excessive table size and lack of smoothness in the \textsf{kyrline} model, \citet{Brenneman2006} created an alternative model for relativistic reflection known as \textsf{kerrdisk} \citep{Brenneman2007}. Their model has a relatively smaller FITS table and a robust interpolation approach. Using a high level of smoothness in the transfer function allows for effective interpolation in \textsf{kerrdisk}. Furthermore, this model employs a distinct methodology, approximating the narrow line of the distant reflection from the accretion disk using a $\delta$-function instead of a Gaussian function. Their model computes a larger portion of the integration using analytic approaches, thereby excluding the emissivity law from the calculated table. However, fitting methods can handle the modeling of the emissivity law. This model effectively reduces the table size to a fraction of the \textsf{kyrline} FITS table. Nevertheless, the relativistic emission produced by \textsf{kerrdisk} appears less smooth than those made by \textsf{kyrline}, with some noticeable spikes in the red wing of the relativistic line. However, data accumulated with the spectral resolutions of detectors aboard the \textit{XMM-Newton} and \textit{Suzaku} telescopes could not distinguish these spikes. The \textsf{kyrline} model has been used to conduct an \textit{XMM-Newton} survey on a sample of radio-quiet Type 1 AGNs \citep{deLaCallePerez2010}. Unlike the fixed-spin models (\textsf{laor} and \textsf{diskline}), \textsf{kyrline} and \textsf{kerrdisk}, which can relativistically be convolved with a reflection model (e.g., \textsf{reflionx}) using the \textsf{kyconv} and \textsf{kerrconv} models, respectively, are more accurate in producing the shape of relativistic emission for any positive spin rates. 
However, a black hole spinning in retrograde relative to the accretion disk ($-1<a<0$) could not have its spin constrained by the \textsf{kyrline} and \textsf{kerrdisk} models.

Since 2010, several spectral models have been developed for relativistic blurred emission (see Table~\ref{ref:rel:model}) that incorporate both positive and negative spin values, enabling the measurement of the black hole's spin in both the prograde and retrograde directions with respect to the accretion disk. To accommodate the full spin range ($-0.998 \leqslant a \leqslant + 0.998$), \citet{Dauser2010} has created the \textsf{relline} model, together with \textsf{relline\_lp}, featuring the `lamp post' geometry \citep[for detail see][]{Dauser2010a}, using FITS tables for Cunningham's photon transfer function \citep{Cunningham1975} pre-calculated with a customized version of the F77 program \textsf{photon\_transferfct}\footnote{\url{https://pisrv1.am14.uni-tuebingen.de/~speith/misc.html}} \citep[also called \textsf{spx};][]{Speith1993,Speith1995} (a gravitationally-distorted appearance of an accretion disk made with a modified version of this program is shown in the bottom-left panel of Fig.~\ref{fig:1}). The \textsf{relline} model employs Green's functions to calculate the radiated radiation for arbitrary angular and radial variations, as well as robust interpolation techniques that lead to a decrease in pre-calculated tabulated data. Moreover, relativistic line profiles calculated for a hard X-ray source located on the rotational axis at a height ($h$) above the black hole, i.e., the `lamp post' geometry, are provided in the \textsf{relline\_lp} model \citep{Dauser2013}. Both the models includes prescriptions for the limb-darkening/-brightening laws. Their corresponding convolution models, \textsf{relconv} and \textsf{relconv\_lp}, are able to convolve the reflection continuum created by a reflection model such as \textsf{reflionx} \citep{Ross1999,Ross2005,Ross2007} and \textsf{xillver} \citep{Garcia2010a,Garcia2010,Garcia2011,Garcia2013}.
As this simple combination of the relativistic convolution model (\textsf{relconv}) and the reflection model (\textsf{xillver}) can lead to inconsistent results, \citet{Garcia2014} made a self-consistent implementation of the spectrum reflected from the disk irradiated by an ionizing source and relativistically blurred emission in a new model called \textsf{relxill}, as well as an additional new model called \textsf{relxill\_lp} for a lamp-post geometry.\footnote{\url{https://www.sternwarte.uni-erlangen.de/~dauser/research/relxill/}} These models incorporated angle-dependent reflection tables of \textsf{xillver}\footnote{\url{https://sites.srl.caltech.edu/~javier/xillver/}} into the relativistic blurring calculations, which exhibits like behavior to a convolution of \textsf{xillver} and \textsf{relconv} (or \textsf{relconv\_lp}), albeit with self-consistently calculated X-ray reflection \citep{Dauser2014a}. Subsequently,  \citet{Dauser2014,Dauser2016} further extended them to include the reflection fraction parameter ($f_{\rm refl,rel}$), a flux ratio between the direct and reflected radiation that depends on the geometry and location of the radiation source. In the updated model \textsf{relxillCp} \citep{Dauser2016}, the primary source is made by a thermally Comptonized continuum model \citep[\textsf{nthcomp};][]{Zdziarski1996,Zycki1999} and offers a free parameter for the disk density ranging from $10^{15}$ to $10^{20}$\,cm$^{-3}$, whereas the previous model \textsf{relxill} uses a power law with a high-energy exponential cutoff (\textsf{zcutoffpl}) and assumes a disk density of $10^{15}$\,cm$^{-3}$. Recently, the inclusion of returning radiation is implemented in the latest model of \textsf{relxilllpCp} by \citet{Dauser2022}, featuring the `lamp post' geometry with a thermally comptonized continuum as the primary source and an unrestricted density parameter ($10^{15}$--$10^{20}$\,cm$^{-3}$). Additionally, \citet{Garcia2022} also present the model \textsf{relxillNS} featuring a black body spectrum, specifically tailored to accommodate the reflection from the disk around an accreting neutron star. 
The relativistic X-ray reflection models provided by the \textsf{relxill} package were also extended to describe the reflection spectrum in the Johannsen metric \citep{Johannsen2013}, referred to as \textsf{relline\_nk} \citep{Bambi2017,Abdikamalov2019,Abdikamalov2020,Abdikamalov2021a,Abdikamalov2021},\footnote{\url{https://www.tat.physik.uni-tuebingen.de/~nampalliwar/relxill_nk/}}\,\footnote{\url{https://github.com/ABHModels/relxill_nk}, doi:\href{https://doi.org/10.5281/zenodo.13906295}{10.5281/zenodo.13906295}} whose model names contain ``nk'' at the end  (e.g., \textsf{relxill\_nk} and \textsf{relxillCp\_nk} stand for non-Kerr spacetimes) to distinguish them from those in the Kerr metric (see Table~\ref{ref:rel:model}). These models allow to validate the Kerr metric through the deformation parameters  $\alpha_{13}$, $\alpha_{22}$, or $\epsilon_{3}$ in the Johannsen metric, where $\alpha_{13}=\alpha_{22}=\epsilon_{3}=0$ restores the Kerr metric \citep[for non-Kerr metrics, see review by][]{Bambi2017a}.

\begin{table*}
\begin{center}
\caption{A list of spectral models developed for relativistically broadened emission of the accretion disk.
\label{ref:rel:model}}
\footnotesize
\begin{tabular}{llll}
\noalign{\smallskip}
\hline\noalign{\smallskip}
\multicolumn{4}{c}{Relativistic Broad Line}\\
\noalign{\smallskip}
\hline
\noalign{\smallskip}
Model   &  Parameters\,$^{\mathrm{\bf a}}$ & Convolution & References\\
\noalign{\smallskip}\hline
\noalign{\smallskip}
\textsf{diskline}   & $a=0$, $i$, $\Gamma$, $R_{\rm in}$, $R_{\rm out}$  & \textsf{rdblur}  &   \citet{Fabian1989} \\
\noalign{\smallskip}
\textsf{laor}   & $a=0.998$, $i$, $q$, $R_{\rm in}$, $R_{\rm out}$  & \textsf{kdblur}  &   \citet{Laor1991} \\
\noalign{\smallskip}
\textsf{kerrspec}   & $a=0.001$, $0.9981$, $i$, $\Gamma$, $R_{\rm in}$, $R_{\rm out}$, $h$  & -- &   \citet{Martocchia2000} \\
\noalign{\smallskip}
\textsf{kyrline/ky}   & $a$ ($\geq0$), $i$, $R_{\rm in}$, $R_{\rm out}$, $R_{\rm br}$, $q_{\rm in}$, $q_{\rm out}$, \textsf{limb}  & \textsf{kyconv} &   \citet{Dovciak2004,Dovciak2022} \\ 
\noalign{\smallskip}
\textsf{kerrdisk}   & $a$ ($\geq0$), $i$, $R_{\rm in}$, $R_{\rm out}$, $R_{\rm br}$, $q_{\rm in}$, $q_{\rm out}$  & \textsf{kerrconv} &   \citet{Brenneman2006} \\ 
\noalign{\smallskip}
\textsf{relline}   & $a$, $i$, $R_{\rm in}$, $R_{\rm out}$, $R_{\rm br}$, $q_{\rm in}$, $q_{\rm out}$, \textsf{limb}  & \textsf{relconv} &   \citet{Dauser2010} \\ 
\noalign{\smallskip}
\textsf{relline\_lp}   & $a$, $i$, $\Gamma$, $R_{\rm in}$, $R_{\rm out}$, $R_{\rm br}$, $h$, \textsf{limb}  & \textsf{relconv\_lp} &   \citet{Dauser2013} \\ 
\noalign{\smallskip}
\textsf{relxill}\,$^{\mathrm{\bf b}}$   & $a$, $i$, $\Gamma$, $R_{\rm in}$, $R_{\rm out}$, $R_{\rm br}$, $q_{\rm in}$, $q_{\rm out}$, & -- &   \citet{Garcia2014}, \\ 
 & $\log \xi$, $A_{\rm Fe}$, $f_{\rm refl,rel}$, $E_{\rm cut}$ &  &  \citet{Dauser2014,Dauser2016} \\ 
\noalign{\smallskip}
\textsf{relxilllp}\,$^{\mathrm{\bf b}}$   & $a$, $i$, $\Gamma$, $R_{\rm in}$, $R_{\rm out}$, $h$, $\beta$, $\log \xi$, $A_{\rm Fe}$, & -- &   \citet{Garcia2014}, \\ 
 & $f_{\rm refl,rel}$, $E_{\rm cut}$ &  &    \citet{Dauser2014,Dauser2016} \\ 
\noalign{\smallskip}
\textsf{relxillCp}\,$^{\mathrm{\bf c}}$   & $a$, $i$, $\Gamma$, $R_{\rm in}$, $R_{\rm out}$, $R_{\rm br}$, $q_{\rm in}$, $q_{\rm out}$, & -- &   \citet{Garcia2014}, \\ 
 & $\log \xi$, $A_{\rm Fe}$, $f_{\rm refl,rel}$, $kT_{\rm e}$, $\log n$ &  &  \citet{Dauser2014,Dauser2016} \\ 
\noalign{\smallskip}
\textsf{relxilllpCp}\,$^{\mathrm{\bf c}}$   & $a$, $i$, $\Gamma$, $R_{\rm in}$, $R_{\rm out}$, $h$, $\beta$, $\log \xi$, $A_{\rm Fe}$, & -- &   \citet{Garcia2014}, \\ 
 & $f_{\rm refl,rel}$, $kT_{\rm e}$, $\log n$ &  &    \citet{Dauser2014,Dauser2016} \\ 
\noalign{\smallskip}
\textsf{relline\_nk}\,$^{\mathrm{\bf d}}$   & $a$, $i$, $R_{\rm in}$, $R_{\rm out}$, $R_{\rm br1}$, $R_{\rm br2}$, $q_{1}$, $q_{2}$, $q_{3}$,  & \textsf{relconv\_nk} &   \citet{Bambi2017}, \\ 
 & \textsf{limb}, $\dot{m}$, $\alpha_{13}$ / $\alpha_{22}$ / $\epsilon_{3}$  &  &    \citet{Abdikamalov2019,Abdikamalov2020} \\
\noalign{\smallskip}
\textsf{rellinelp\_nk}\,$^{\mathrm{\bf d}}$   & $a$, $i$, $\Gamma$, $R_{\rm in}$, $R_{\rm out}$, $R_{\rm br}$, $h$, \textsf{limb}, $\dot{m}$,  & \textsf{relconvlp\_nk} &   \citet{Bambi2017} \\ 
 & $\alpha_{13}$ / $\alpha_{22}$ / $\epsilon_{3}$  &  &    \citet{Abdikamalov2019,Abdikamalov2020} \\
\noalign{\smallskip}
\textsf{relxill\_nk}\,$^{\mathrm{\bf d}}$  & $a$, $i$, $\Gamma$, $R_{\rm in}$, $R_{\rm out}$, $R_{\rm br}$, $R_{\rm br}$, $q_{1}$, $q_{2}$, $q_{3}$, & -- &   \citet{Bambi2017}, \\ 
 & $\log \xi$, $A_{\rm Fe}$, $f_{\rm refl,rel}$, $E_{\rm cut}$, $\dot{m}$, $\alpha_{13}$ / $\alpha_{22}$ / $\epsilon_{3}$  &  &  \citet{Abdikamalov2019,Abdikamalov2020} \\ 
\noalign{\smallskip}
\textsf{relxilllp\_nk}\,$^{\mathrm{\bf d}}$   & $a$, $i$, $\Gamma$, $R_{\rm in}$, $R_{\rm out}$, $h$, $\log \xi$, $A_{\rm Fe}$, $f_{\rm refl,rel}$, & -- &   \citet{Bambi2017}, \\ 
 & $E_{\rm cut}$, $\dot{m}$, $\alpha_{13}$ / $\alpha_{22}$ / $\epsilon_{3}$ &  &    \citet{Abdikamalov2019,Abdikamalov2020} \\ 
\noalign{\smallskip}
\textsf{relxillCp\_nk}\,$^{\mathrm{\bf d}}$   & $a$, $i$, $\Gamma$, $R_{\rm in}$, $R_{\rm out}$, $R_{\rm br1}$, $R_{\rm br2}$, $q_{\rm 1}$, $q_{\rm 2}$, $q_{\rm 3}$, $\log \xi$, & -- &   \citet{Bambi2017}, \\ 
 & $A_{\rm Fe}$, $f_{\rm refl,rel}$, $kT_{\rm e}$, $\log n$, $\dot{m}$, $\alpha_{13}$ / $\alpha_{22}$ / $\epsilon_{3}$ &  &  \citet{Abdikamalov2019,Abdikamalov2020} \\ 
\noalign{\smallskip}
\textsf{relxilllpCp\_nk}\,$^{\mathrm{\bf d}}$   & $a$, $i$, $\Gamma$, $R_{\rm in}$, $R_{\rm out}$, $h$, $\log \xi$, $A_{\rm Fe}$, $f_{\rm refl,rel}$, & -- &   \citet{Bambi2017}, \\ 
 & $kT_{\rm e}$, $\log n$, $\dot{m}$, $\alpha_{13}$ / $\alpha_{22}$ / $\epsilon_{3}$ &  &    \citet{Abdikamalov2019,Abdikamalov2020} \\ 
\noalign{\smallskip}
\textsf{reflkerr}   & $a$, $i$, $\tau$ / $\Gamma$, $R_{\rm in}$, $R_{\rm out}$, $R_{\rm br}$, $q_{\rm in}$, $q_{\rm out}$, $\log \xi$, $A_{\rm Fe}$, & -- &   \citet{Niedzwiecki2008}, \\ 
 & $f_{\rm refl,rel}$, $kT_{\rm e}$, $kT_{\rm bb}$, \textsf{geom} &  & \citet{Niedzwiecki2016,Niedzwiecki2019} \\ 
\noalign{\smallskip}
\textsf{reflkerr\_lp}   & $a$, $i$, $\tau$ / $\Gamma$, $R_{\rm in}$, $R_{\rm out}$, $h$, $\delta$, $\log \xi$, $A_{\rm Fe}$, $f_{\rm refl,rel}$, & -- &   \citet{Niedzwiecki2008}, \\ 
 & $kT_{\rm e}$, $kT_{\rm bb}$, \textsf{geom} &  & \citet{Niedzwiecki2016,Niedzwiecki2019} \\ 
\noalign{\smallskip}
\textsf{reltrans}   & $a$, $i$, $\Gamma$, $R_{\rm in}$, $R_{\rm out}$, $h$, $\log \xi$, $A_{\rm Fe}$, $E_{\rm cut}$, $N_{\rm H}$, & -- &   \citet{Ingram2019}, \\ 
 & $1/\mathcal{B}$, $M_{\rm BH}$, $\nu_{\rm min}$, $\nu_{\rm max}$, $\phi_{\rm A}$, \textsf{ReIm} &  &  \citet{Mastroserio2021} \\ 
\noalign{\smallskip}
\textsf{reltransCp}   & $a$, $i$, $\Gamma$, $R_{\rm in}$, $R_{\rm out}$, $h$, $\log \xi$, $A_{\rm Fe}$, $kT_{\rm e}$, $N_{\rm H}$, & -- &   \citet{Ingram2019}, \\ 
 & $1/\mathcal{B}$, $M_{\rm BH}$, $\nu_{\rm min}$, $\nu_{\rm max}$, $\phi_{\rm A}$, \textsf{ReIm} &  &  \citet{Mastroserio2021} \\
\noalign{\smallskip}
\hline\noalign{\smallskip}
\noalign{\smallskip}
\end{tabular}
\end{center}
\begin{tablenotes}
\footnotesize
\item[1]\textbf{Notes.} $^{\mathrm{\bf a}}$ Parameters of relativistic broad line models are as follows:
dimensionless black-hole spin parameter ($a$), disk inclination angle relative to the line of sight ($i$), power-law index ($\Gamma$), inner radius ($R_{\rm in}$), outer radius ($R_{\rm out}$), break radius ($R_{\rm br}$),
coronal emissivity law indexes ($q$ as $r^{-q}$ between $R_{\rm in}$ and $R_{\rm out}$; $q_{\rm in}$ as $r^{-q_{\rm in}}$ between $R_{\rm in}$ and $R_{\rm br}$; and $q_{\rm out}$ as $r^{-q_{\rm out}}$ between $R_{\rm br}$ and $R_{\rm out}$), 
height of the primary source above the black hole ($h$), velocity of the primary source relative to the speed of light ($\beta$), \textsf{limb} describes limb-darkening/-brightening law (0: isotropic emission, 1: Laor's limb-darkening $1+2.06\mu$, and 2: Haardt's limb-brightening $\ln [1+1/\mu]$), ionization parameter of the accretion disk ($\log \xi$), iron abundance relative to solar ($A_{\rm Fe}$), reflection fraction parameter ($f_{\rm refl,rel}$), high energy cutoff ($E_{\rm cut}$) of the primary source described by \textsf{cutoffpl}, the electron temperature ($kT_{\rm e}$) in the corona described by \textsf{nthcomp}, logarithmic number density of the innermost accretion disk ($\log n$), break radii ($R_{\rm br1}$ and $R_{\rm br2}$) for non-Kerr spacetimes, coronal emissivity law indexes in non-Kerr spacetimes ($q_{\rm 1}$ between $R_{\rm in}$ and $R_{\rm br1}$; $q_{\rm 2}$ between $R_{\rm br1}$ and $R_{\rm br2}$; and $q_{\rm 2}$ between $R_{\rm br2}$ and $R_{\rm out}$), hydrogen column density of the line-of-sight material ($N_{\rm H}$), boosting factor of the reflection spectrum ($1/\mathcal{B}$), black hole mass ($M_{\rm BH}$), frequency range of the transfer function ($\nu_{\rm min}$ and $\nu_{\rm max}$), phase normalization ($\phi_{\rm A}$), cross-spectrum modes (\textsf{ReIm}\,$=1, 2, 3, 4, 5, $ or $6$), blackbody soft-excess temperature ($kT_{\rm bb}$), Thomson optical depth $\tau(\Gamma,kT_{\rm e})$ yielding $\Gamma$, bottom-lamp attenuation ($0<\delta<1$), and geometry (\textsf{geom}$=$\,$-5, -4, 0, 4,$ or $5$) defined similar to the same parameter in \textsf{compPS} \citep{Poutanen1996}.  
\newline
$^{\mathrm{\bf b}}$ The model \textsf{relxill} is a combination of \textsf{relconv}, \textsf{xillver}, and \textsf{cutoffpl}. The model \textsf{relxilllp} is a mixture of \textsf{relconv\_lp}, \textsf{xillver}, and \textsf{cutoffpl}. The number density is fixed  ($\log n=15$) in the models \textsf{relxill} and \textsf{relxilllp}.
\newline
$^{\mathrm{\bf c}}$ The models \textsf{relxillCp} and \textsf{relxilllpCp} are, respectively, similar to \textsf{relxill} and \textsf{relxilllp}, but they use \textsf{nthcomp} instead of \textsf{cutoffattenuation of the pl}, as well as a free parameter for the number density ($\log n=15$--$20$).
\newline
$^{\mathrm{\bf d}}$ ``\textsf{nk}'' at the end of the spectral models stands for non-Kerr spacetimes, which are describe by the deformation parameters  $\alpha_{13}$, $\alpha_{22}$, or $\epsilon_{3}$ in the Johannsen metric \citep{Johannsen2013}, as well as the thickness of the accretion disk described by $\dot{m}$ (0: infinitesimally-thin, 1: 5\%, 2: 10\%, 3: 20\%, 4: 30\% of the Eddington accretion rate).
\end{tablenotes}
\end{table*}

X-ray time-resolved observations of AGNs have shown \textit{variability} in the relativistically blurred reflection \citep[e.g., MCG--6-30-15;][]{Fabian2003,Vaughan2004,Larsson2007,Miller2008} that could be caused by general relativistic effects, particularly light bending near the black hole event horizon. \citet{Niedzwiecki2008} and \citet{Niedzwiecki2010} investigated variability patterns of the red wing in X-ray reflection of the AGN in the Seyfert 1 galaxy MCG--6-30-15 using a detailed light-bending model \citep{Miniutti2004}, which led to the development of the spectral model \textsf{reflkerr} and its corresponding lamp-post model \textsf{reflkerr\_lp} \citep{Niedzwiecki2016,Niedzwiecki2019}.\footnote{\url{https://users.camk.edu.pl/mitsza/reflkerr/}} In particular, \citet{Niedzwiecki2016} identified some inconsistencies between \textsf{reflkerr} and \textsf{relxilllp} owing to the neglect of the general-relativistic redshift of the direct coronal radiation in \textsf{relxilllp}, though they found that the \textsf{relxilllp} model still produces acceptable results in weak-gravity in the energies below 80\,keV. Moreover, the lamp-post model \textsf{reflkerr\_lp} developed by \citet{Niedzwiecki2019} demonstrated a departure from \textsf{relxilllp} in the energies above 30\,keV. Another model-family for spectral and timing variability in accreting black holes has been developed \citep{Ingram2019,Mastroserio2021,Mastroserio2022}, named \textsf{reltrans} and \textsf{reltransCp},\footnote{\url{https://adingram.bitbucket.io/reltrans.html}} which calculated the emergent reflection spectrum using \textsf{xillver} (or \textsf{xillverCp} in the case of \textsf{reltransCp}). The \textsf{reltrans} model considers all the general-relativistic effects to calculate the time delays and energy changes that occur when X-ray photons from the corona reflect from the accretion disk and scatter towards the observer. The calculations of \textsf{reltrans} incorporate both continuum lags and reverberation lags in a self-consistent manner to produce most of the practical X-ray variability time scales.

\section{Variability in Relativistic Reflection from PCA}
\label{review:pca}

Principal component analysis \citep[PCA;][]{Hotelling1933},\footnote{It was first innovated by \citet{Pearson1901} in the context of principal axes of ellipsoids in geometry, but it was independently developed and called \textit{the method of principal components} by \citet{Hotelling1933} for statistical analysis.} also referred to as the `Hotelling transform', is a well-known method in multivariate statistics relying on eigenvalues and eigenvectors \citep[see review by][]{Jolliffe2016} that has been extensively discussed in detail in the literature \citep[e.g.,][]{Mardia1979,Jolliffe2002,Izenman2008,Rencher2012}.
It bears a close relation to the `Kosambi--Karhunen--Lo\`{e}ve transform' \citep{Kosambi1943,Karhunen1947,Loeve1948} in probability theory, and is among three classical techniques in multivariate analysis to determine the principal dimensions of large data, along with independent component analysis \citep[ICA;][]{Herault1984,Herault1985,Herault1986} and non-negative matrix factorization \citep[NMF;][]{Lee1999,Lee2000}. PCA can be employed to separate various characteristics that are mostly responsible for complex variations in large data in astronomy \citep[e.g.,][]{Wall2012,Ivezic2020} 
as well as to simplify complex data for machine learning approaches \citep[e.g.,][]{Bishop2006,Mueller2016,Witten2017,Geron2019}. This is implemented by reducing the number of available data into a group of independent PCA components, which then provide information about the different levels of their contributions to the complexity of the entire data. Astronomers have extensively employed it as a practical multivariate method. The early application of this technique in astronomy \citep[see review by][]{Francis1999} can be traced back to some studies on spectral analyses of stars \citep{Deeming1964,Whitney1983}, galaxies \citep{Faber1973,Bujarrabal1981,Efstathiou1984}, and quasars \citep{Mittaz1990,Francis1992,Boroson1992}. 
This approach was also employed for imaging analysis of the interstellar medium \citep{Heyer1997,Brunt2009}. It was later used for X-ray binaries \citep[e.g.,][]{Malzac2006,Koljonen2013,Koljonen2015} and blazars \citep{Gallant2018}, and more recently for X-ray variability in symbiotic stars \citep{Danehkar2024a} and starburst regions \citep{Danehkar2024}. Especially, it has extensively been leveraged for X-ray data analysis of AGNs in Seyfert 1 galaxies \citep[e.g.,][]{Vaughan2004,Miller2008,Parker2014,Gallo2015}.

PCA can decompose time-resolved spectroscopic data into groups of PCA components and eigenvectors, yielding eigenvalues in the process. Normalized eigenvalues can yield the contribution of each eigenvector to the temporal evolution of the whole data over time. Each decomposed PCA component and eigenvector can be referred to as a principal spectrum with its corresponding light curve. The process of conducting PCA requires performing the decomposition of a matrix into its eigenvectors and eigenvalues. To analyze variability of a source in astronomy, this data matrix for PCA contains a set of spectroscopic data collected at $n$ time intervals, each binned into $m$ spectral channels.
Let consider a rectangular ($m \times n$) matrix $\mathbf{X}$ consisting of $n$ rows by $m$ columns, one can determine the principal components of $\mathbf{X}$ through the following three methods:
\begin{itemize}
\item \textbf{Singular Value Decomposition.} The most common approach to obtaining the PCA components is the singular value decomposition \citep[SVD;][]{Beltrami1873,Jordan1874a,Jordan1874b,Sylvester1889a,Sylvester1889b,Sylvester1889c}. The SVD of $\mathbf{X}$ is performed as follows:
\begin{equation}
\mathbf{X} =  \mathbf{U}\mathbf{\Sigma}\mathbf{V}^{\intercal},  
\label{eq_2_1}
\end{equation}
where $\mathbf{U}$ is a square matrix of order $m$ containing \textit{principal components} (spectra in time-resolved spectroscopic data), $\mathbf{\Sigma}$ is a rectangular diagonal matrix ($m \times n$) containing square roots of \textit{eigenvalues} in its diagonal, i.e, $\mathbf{\Lambda}=\mathbf{\Sigma}^{\intercal}\mathbf{\Sigma}=\mathrm{diag}(\lambda_1,\ldots,\lambda_n)$ (contribution fractions in spectral variability), $\mathbf{V}$ is a square matrix of order $n$ containing \textit{eigenvectors} (light curves in astronomical data).

\item \textbf{Eigendecomposition.} A classical way to determine the PCA components is through the eigenvalue
decomposition \citep[EVD;][]{Cauchy1829,Cauchy1829a}\footnote{For a historical review, see \citet{Hawkins1975}.} of the covariance matrix expressed as $\mathbf{C}_{\mathrm{XX}} =  \mathbf{X}^{\intercal} \mathbf{X}$, which is a square matrix of order $m$. The eigendecomposition of $\mathbf{C}_{\mathrm{XX}}$ is as follows:
\begin{equation}
\mathbf{C}_{\mathrm{XX}} =  \mathbf{V}\mathbf{\Lambda}\mathbf{V}^{\intercal},
\label{eq_2_3}
\end{equation}
which yields eigenvectors ($\mathbf{V}$) and eigenvalues ($\mathbf{\Lambda}$). The principal components ($\mathbf{U}$) are obtained from the decomposed eigenvectors and eigenvalues by considering Eq. (\ref{eq_2_1}), which leads to the following solution:\footnote{Based on the fast that $\mathbf{V}^{\intercal}\mathbf{V}=\mathbf{I}$, where $\mathbf{I}=\mathrm{diag}(1,\ldots,1)$ is the identity matrix of order $n$, so $\mathbf{\Sigma}\mathbf{I} =\mathbf{\Sigma}$ and $\mathbf{\Lambda}\mathbf{I} =\mathbf{\Sigma}$.}
\begin{equation}
\mathbf{X}\mathbf{V} = \mathbf{U}\mathbf{\Sigma}\mathbf{V}^{\intercal}\mathbf{V}=\mathbf{U}\mathbf{\Sigma}.  
\label{eq_2_4}
\end{equation}
Constructing the diagonal matrix $\mathbf{\Sigma}=\mathbf{\Lambda}^{\frac{1}{2}}$ from the eigenvalues ($\mathbf{\Lambda}$) and obtaining the least-squares solution to $\mathbf{X}\mathbf{V}=\mathbf{U}\mathbf{\Sigma}$ lead to the principal components ($\mathbf{U}$) of $\mathbf{X}$.

\item \textbf{QR Decomposition.} Another faster method suitable for high-performance computing, which was proposed by \citet{Sharma2013} to conduct PCA, is performed using QR decomposition \citep{Golub1965}, also known as QR factorization \citep{Golub1996,Trefethen1997}. In this approach, $\mathbf{X}$ is first factorized into an orthogonal matrix $\mathbf{R}$ of $n \times m$ dimensions and an upper triangular square matrix $\mathbf{R}$ of order $m$:
\begin{equation}
\mathbf{X} =  \mathbf{Q}\mathbf{R}.  
\label{eq_2_5}
\end{equation}
Then, the SVD of $\mathbf{R}^{\intercal}$ is obtained:
\begin{equation}
\mathbf{R}^{\intercal} =  \mathbf{\bar{U}}\mathbf{\bar{\Sigma}}\mathbf{\bar{V}}^{\intercal}.
\label{eq_2_6}
\end{equation}
As demonstrated by \citet{Sharma2013}, this leads to the same diagonal matrix and eigenvectors of Eq. (\ref{eq_2_1}), $\mathbf{\Sigma}=\mathbf{\bar{\Sigma}}$ and $\mathbf{V}= \mathbf{\bar{U}}$, while the equivalent principal components are obtained via $\mathbf{U}=\mathbf{Q} \mathbf{\bar{V}}$.
\end{itemize}
For a set of timing spectroscopic data stored in a data matrix, the PCA components ($\mathbf{U}$), the eigenvectors ($\mathbf{V}$), and the eigenvalues ($\mathbf{\Lambda}=\mathbf{\Sigma}^{2}$) decomposed from the data matrix via either SVD, EVD, or QR are the principal spectra, the corresponding light curves, and their contribution fractions, respectively. As shown in Supplementary Material (\ref{review:supp}), these PCA methods can simply implemented with the linear algebra  functions of \textsf{NumPy} in Python, albeit with neither CPU parallelization nor GPU acceleration. Currently, there are two publicly available packages made for PCA in the astronomical community: (1) the SVD-based Python package \textsc{pca}\footnote{\url{https://www.michaelparker.space/pca-code}} \citep{Parker2015} based on the SVD function \citep[][]{Press1997} from \textsf{NumPy} \citep{Harris2020}, and (2) the QR-based library \textsf{qrpca} in Python\footnote{\url{https://github.com/xuquanfeng/qrpca}} and R\footnote{\url{https://github.com/RafaelSdeSouza/qrpca}} \citep{deSouza2022,deSouza2022a} implemented with \textsf{pyTorch} \citep{Paszke2019} and \textsf{Scikit-learn} \citep{Pedregosa2011} in Python, and with \textsf{torch} \citep{Falbel2022} and the built-in \textsf{prcomp} function in R, allowing for seamless GPU acceleration. In particular, the package \textsc{pca} distributed by \citet{Parker2015}   is capable of conducting PCA on X-ray \textit{XMM-Newton} EPIC-pn observations. To perform PCA with this package, it is necessary to generate a set of X-ray spectra sliced at fixed time intervals (e.g., 10 ks) from event data via custom reduction methods \citep[for details, see][]{Danehkar2024a}.

\begin{figure}[h!]
\begin{center}
\includegraphics[width=0.56\textwidth, trim = 0 0 0 0, clip, angle=0]{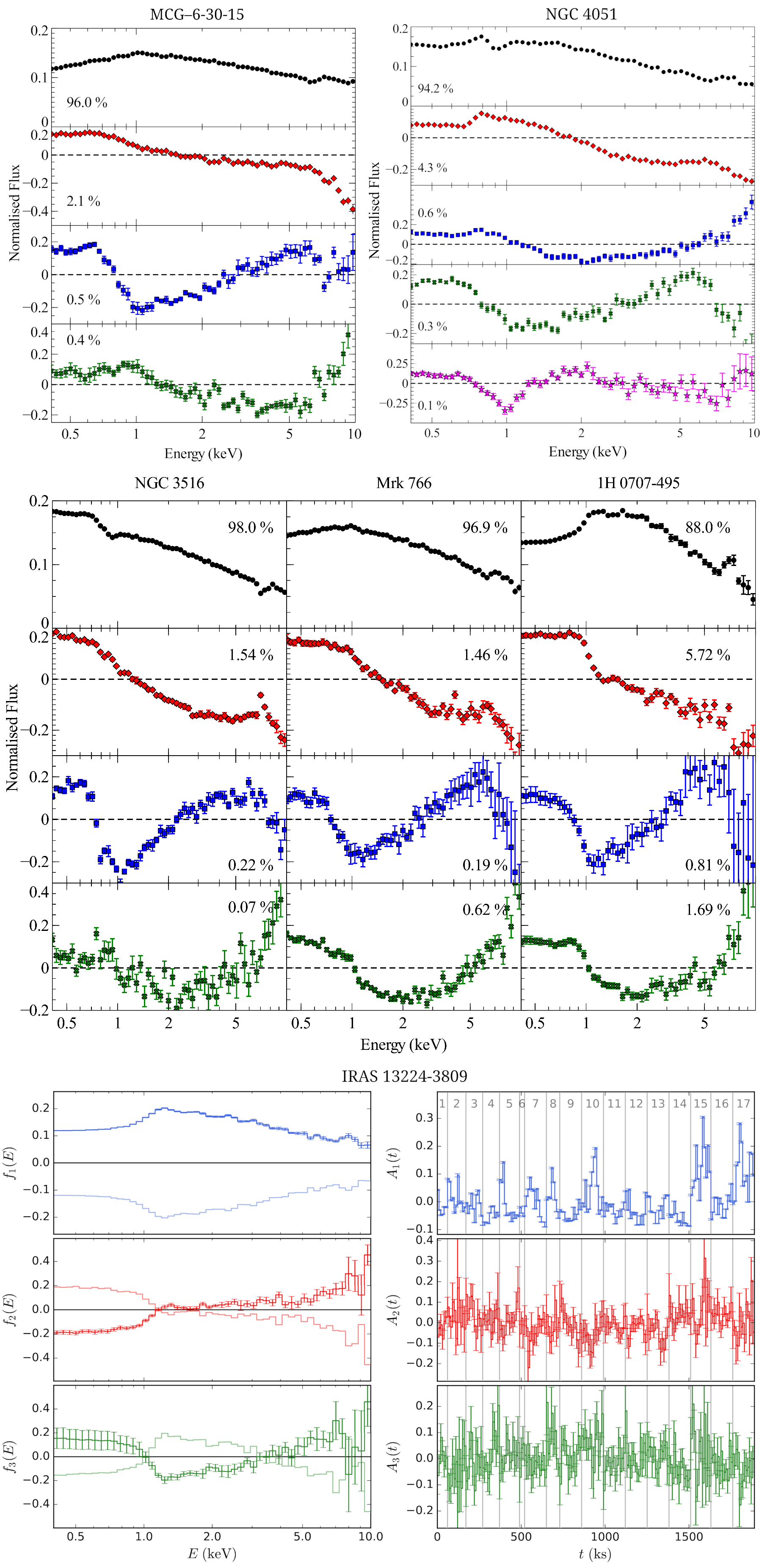}
\end{center}
\caption{PCA spectra found in different AGNs hosted by nearby Seyfert 1 galaxies: MCG\,--6-30-15 \citep{Parker2014a}, NGC\,4051, NGC\,3516, Mrk\,766, and 1H\,0707-495 \citep{Parker2015}, with percentages of variability fractions, as well as PCA spectra $f_{\textrm{1-3}}(E)$ and related light curves $A_{\textrm{1-3}}(t)$ found in  IRAS\,13224-3809 \citep{Parker2017}. The PCA components from the first to the third or/and fourth order, respectively, correspond to variations in the power-law normalization, the power-law spectral index, and the relativistic reflection.
\label{fig:2}
}
\end{figure}

\citet{Vaughan2004} made initial attempts to conduct PCA on X-ray variability in AGN using low-spectral resolution data, suggesting that the X-ray variations in MCG--6-30-15 reported by \citet{Fabian2003} are primarily due to a variable power-law component, with a small partial fraction likely originating from a reflection-dominated component. Later, \citet{Miller2007a} employed SVD for PCA, resulting in the generation of exhaustive principal spectra of Mrk\,766, which \citet{Turner2007} confirmed these spectral variations through time-resolved spectroscopy. Moreover, \citet{Miller2008} investigated the X-ray variability of MCG--6-30-15 using PCA, resulting in similar spectral components (absorbed, varying power-law) in MCG--6-30-15 and Mrk\,766, with a less variable, heavily absorbed component characterizing the relativistically broadened red wing. PCA conducted by \citet{Parker2014a,Parker2014} demonstrated that SVD can successfully separate different spectral components responsible for the X-ray variability in AGNs by exploiting large archival data. In particular, \citet{Parker2014a} discovered that the X-ray variations in MCG--6-30-15 are mostly caused by only three spectral components (see Fig.\,\ref{fig:2}): the normalization factor of the power-law continuum (variability fraction of $\sim 96$\%), the power-law spectral index ($\sim 2.1$\%), and the normalization factor of a relativistically broadened reflection emission ($\sim 0.5$\%). Similarly, PCA by \citet{Parker2015} provided evidence for the slight variability ($\lesssim 0.5$\%) of relativistic reflection in other AGNs hosted by other Seyfert 1 galaxies (NGC\,4051, NGC\,3516, Mrk\,766, and 1H\,0707-495), as seen in Fig.\,\ref{fig:2}. Various spectral analysis approaches, including PCA,  performed by \citet{Gallo2015} also indicated that the variability in the narrow-line Seyfert 1 galaxy, Mrk 335, is mostly caused by changes in the power-law flux and photon index, although small variations in the ionization state of the reflection were found to be necessary. The PCA study of the extreme narrow-line Seyfert 1 galaxy IRAS\,13224-3809 by \citet{Parker2017} also showed three principal spectra: a varying power-law continuum, a slightly variable soft excess, and a less variable broad soft excess being linked to strong reflection (see Fig.\,\ref{fig:2} bottom). In addition, the PCA component associated with a variable power-law continuum contain absorption footprints caused by the relativistic UFO detected by \citet{Parker2017a}. 

As seen in Fig.\,\ref{fig:2}, the third or/and fourth PCA components obtained by \citeauthor{Parker2015} from X-ray observations of five AGNs (MCG\,--6-30-15, NGC\,4051, NGC\,3516, Mrk\,766, and 1H\,0707-495) resemble the relativistically broadened iron emission features shown in Fig.~\ref{fig:1} (top). Their normalized eigenvalues of $\lesssim 0.5$\% imply that they have negligible variations compared to the X-ray variability in the power-low source continuum. This insignificant variability in the relativistic emission is consistent with the fact that the SMBH spin remains constant over the course of the human timescale. Magnetohydrodynamic (MHD) simulations of an black hole accretion disk by \citet{Schnittman2006} revealed that the light curves indeed contain very low levels of variability \citep[see animations by][]{Schnittman2019}. Further MHD simulations by \citet{Schnittman2013} suggested that the noticeable X-ray variability mostly originates from the corona and not the disk. This is in agreement with the results found by \citeauthor{Parker2015}, which show $\gtrsim 90$\% of X-ray variations are due to changes in the power-law continuum, i.e., the corona. 
General-relativistic magnetohydrodynamic (GRMHD) simulations by \citet{Shiokawa2017} also indicated the presence of some flux fluctuations in the emission from the innermost accretion disk due to the fast-moving turbulent formations, as well as some variations in the photon ring with spin-dependent frequencies \citep[see animations by][]{Shiokawa2017a}. Interestingly, the polarimetric light-curve observations of Sgr\,A* have also shown intraday variability in circular polarization \citep{Bower2002} and linear polarization \citep{Marrone2006}, as well as Faraday rotation variability on timescales from hours to months \citep{Marrone2007,Bower2018}. Similarly, the near-infrared GRAVITY-Very Large Telescope Interferometer (VLTI) observations exhibited that the polarization loop in Sgr\,A* is regularly changing clockwise over $\sim 30$ min, indicating a closed, loop motion with the speed of $0.3c$ \citep{GRAVITYCollaboration2018}. The observed polarization variations in Sgr\,A* were in line with predictions from general-relativistic ray-tracing models of slightly tilted accretion flows in the presence of powerful magnetic fields \citep{GRAVITYCollaboration2020}. More recently, the polarimetric Event Horizon Telescope (EHT) imaging observations of the SMBHs in M87 and SgrA* confirmed rapid (intrahour) intrinsic variations in near-horizon accretion flows and polarized rings, which were attributed to spiraling polarization structures based on the results from GRMHD simulations \citep{EventHorizonTelescopeCollaboration2021a,EventHorizonTelescopeCollaboration2021,EventHorizonTelescopeCollaboration2023,EventHorizonTelescopeCollaboration2024,EventHorizonTelescopeCollaboration2024a}. 
Therefore, the recent polarization GRAVITY-VLTI and EHT imaging observations of two nearby SMBHs (Sgr\,A* and M87), together with numerical simulations, imply that small variations in the relativistically broadened iron emission revealed by PCA could be associated with intrahour intrinsic variations and varying spiraling polarization features in near-horizon accretion flows and photon rings.


\section{Future Perspective: Machine Learning}

SVD and PCA decomposition closely relate to the optimal solution for neural networks in auto-association mode \citep{Bourlard1988,Baldi1989}. As discussed by \citet{Hertz1991} in the context of unsupervised Hebbian learning, PCA can be used for dimensionality reduction of large data before proceeding with machine learning algorithms, such as artificial neural networks (ANNs). PCA can indeed alleviate the `curse of dimensionality' \citep[coined by][]{Bellman1957,Bellman1961}, also known as the `Hughes phenomenon' \citep{Hughes1968} or `peaking phenomenon' \citep{Trunk1979}, which often arises when searching for patterns in unknown large data. 
It has been extensively demonstrated in the literature that PCA can be utilized as a pre-processing step to simplify complex data prior to machine learning \citep[e.g.,][]{Bishop2006}, data mining \citep[][]{Witten2017}, and deep learning \citep[][]{Goodfellow2017}. Recently, \citet{Ivezic2020} also discussed in detail the applications of PCA, ICA, and NMF in dimensionality reduction for data mining and machine learning in astronomy. 

Using PCA for the pre-processing of astronomical data enables a significant reduction in dimensionality and complexity of data, leading to an improvement in machine learning performance. The use of PCA to reduce the dimensionality of the data for training ANNs can be traced back to earlier efforts on the classification of galaxy spectra \citep{Folkes1996,Lahav1996} and stellar spectra \citep{Bailer-Jones1998,Singh1998}. Later, \citet{Zhang2003} applied PCA to the multiwavelength data of AGNs, stars, and normal galaxies in order to reduce the dimensionality of the parameter space for support vector machines (SVM) and learning vector quantization (LVQ), two supervised classification algorithms in machine learning, resulting in the classification of stars, AGNs, and normal galaxies. PCA also reduced the complexity of image data for the morphological classification of galaxies with an ANN \citep{delaCalleja2004}. Moreover, \citet{Bu2015} deployed PCA to pre-assemble stellar atmospheric parameters from spectra for Gaussian process regression (GPR) and then compared the results of GPR with those from ANNs, kernel regression (KR), and support-vector regression (SVR). \citet{Kuntzer2016} also conducted stellar classification from single-band images using pre-processed data from PCA to train ANNs to determine the spectral type. More recently, we see the application of PCA to construct input data for ANNs in stellar population synthesis modeling \citep{Alsing2020}, finding thermal components in X-ray spectra of the Perseus cluster \citep{Rhea2020}, and finally X-ray spectral analysis of AGN \citep{Parker2022}.

The avenue of automated spectral analysis with machine learning algorithms has not yet been fully explored for constraining the relativistically broadened iron emission in AGN, mostly because of the complicated variability seen in the X-rays over the course of observations. X-ray observations of AGNs have shown some X-ray changes in power-law continua, which were ascribed to so-called transient obscuration events caused by eclipsing material near the primary source, such as NGC\,3783 \citep{Mehdipour2017}, NGC\,3227 \citep{Turner2018}, and Mrk\,335 \citep{Longinotti2019,Parker2019}, or flaring variations in the corona in the innermost central regions, e.g., PDS\,456 \citep{Matzeu2017a,Reeves2021} and NGC\,3516 \citep{Mehdipour2022}. This kind of change in X-rays over time, along with a relatively large number of parameters in relativistic reflection models (see Table~\ref{ref:rel:model}), makes it much more complicated for machine learning algorithms to automatically determine the spins of SMBHs from the archival X-ray data. Nevertheless, as seen in Fig.\,\ref{fig:2}, the dimensionality reduction offered by PCA can avoid the curse of dimensionality in the X-ray data of AGNs. In the future, we will be able to use machine learning to automatically conduct the spin analysis of SMBHs in AGNs thanks to the principal spectra of relativistic reflection disentangled by PCA from X-ray observations.

\section*{Author Contributions}


The author confirms being the sole contributor of this work and has approved it for publication.

\section*{Funding}


The author acknowledges financial support from the National Aeronautics and Space Administration (NASA) for an Astrophysics Data Analysis Program grant under no. 80NSSC22K0626.

\section*{Conflict of Interest Statement}

The author declares that the research was conducted in the absence of any commercial or financial relationships that could be construed as a potential conflict of interest.

\section*{Acknowledgments}

The author would like to express his gratitude for the invitation to speak at the `Frontiers in Astronomy and Space Sciences: A Decade of Discovery and Advancement, 10th Anniversary Conference,' as well as to the editor who requested a concise review of that presentation. The author thanks Michael Parker for permission to use figures from his publications and useful discussions; Javier Garc{\'\i}a, Thomas Dauser, and Laura Brenneman for permission to use figures from their publications; and the reviewer for careful reading of the manuscript and constructive comments.


\section*{Supplemental Data}

The Supplementary Material for this manuscript can be found online here: \ref{review:supp}.

\newpage

\appendix
\renewcommand{\thesection}{\Alph{section}}

\section{Supplementary Material}
\label{review:supp}

\subsection{Three PCA methods implemented with NumPy in Python}

\noindent We can obtain the PCA components ($\mathbf{U}$), the eigenvectors ($\mathbf{V}$), and the eigenvalues ($\mathbf{\Lambda}=\mathbf{S}^{2}$) of a $m \times n$ matrix $\mathbf{X}$ via the following three methods with the \textsf{NumPy} linear algebra. Let consider $\mathbf{X}$ defined as:
\begin{lstlisting}
import numpy as np
np.random.seed(50)
n = 3; m = 10
X = np.random.random((m , n))
\end{lstlisting}

\noindent \textbf{Singular Value Decomposition (SVD).} The SVD of $\mathbf{X} =  \mathbf{U}\mathbf{S}\mathbf{V}^{\intercal}$ can be conducted as follows:
\begin{lstlisting}
U, S, VT = np.linalg.svd(X, full_matrices=False)
V = VT.T
eigenVectors = V
eigenValues = S**2
\end{lstlisting}

\noindent \textbf{Eigendecomposition.} The eigenvalue decomposition (EVD) of the covariance matrix $\mathbf{C}=  \mathbf{X}^{\intercal} \mathbf{X} =  \mathbf{V}\mathbf{\Lambda}\mathbf{V}^{\intercal}$ and the least-squares solution ($\mathbf{U}$) to $\mathbf{X}\mathbf{V}=\mathbf{U}\mathbf{S}$ can be computed as follows:
\begin{lstlisting}
C = np.dot(X.T, X)
eigenValues, eigenVectors = np.linalg.eigh(C)
idx_sort = eigenValues.argsort()[::-1]   
eigenValues = eigenValues[idx_sort]
eigenVectors = eigenVectors[:,idx_sort]
S = eigenValues**0.5
V = eigenVectors
XV = np.dot(X,V)
UT, resid, rank, sv = np.linalg.lstsq(np.diag(S),XV.T, rcond=None)
U = UT.T
\end{lstlisting}

\noindent \textbf{QR Decomposition.} The QR decomposition of $\mathbf{X}= \mathbf{Q}\mathbf{R}$ and the SVD of $\mathbf{R}^{\intercal}= \mathbf{\bar{U}}\mathbf{\bar{S}}\mathbf{\bar{V}}^{\intercal}$ can also be implemented using the \textsf{NumPy} functions, leading to $\mathbf{S}=\mathbf{\bar{S}}$, $\mathbf{V}= \mathbf{\bar{U}}$, and $\mathbf{U}=\mathbf{Q} \mathbf{\bar{V}}$:
\begin{lstlisting}
Q, R =  np.linalg.qr(X)
U_bar, S_bar, VT_bar = np.linalg.svd(R.T)
V_bar = VT_bar.T
U = np.dot(Q, V_bar)
V = U_bar
S = S_bar
eigenVectors = V
eigenValues = S**2
\end{lstlisting}
These different methods can sometimes result in a sign difference in one PCA component and its eigenvector compared to each other.

\end{document}